\documentclass[
    reprint,
    aps,
    prl,
    superscriptaddress,
    amsmath,
    amssymb,
    longbibliography,
    nofootinbib
]{revtex4-2}

\usepackage{mathtools}
\usepackage{tabularx}
\usepackage{array}
\usepackage{graphicx}
\usepackage{xcolor}
\usepackage[normalem]{ulem}
\usepackage{physics}
\usepackage{amsfonts}
\usepackage{float} 

\PassOptionsToPackage{%
  colorlinks=true,
  linkcolor=blue,
  urlcolor=blue,
  citecolor=blue,
  anchorcolor=blue%
}{hyperref}

\usepackage{orcidlink}
\usepackage{comment}
\usepackage{ifthen}
\usepackage{enumitem}

\usepackage{pgfplots}
\usepackage{tikz}
\usepackage{quantikz}
\usetikzlibrary{calc}
\usetikzlibrary{intersections,shapes.arrows}
\usepackage{tikz-3dplot}
\pgfplotsset{compat=1.17}
\tdplotsetmaincoords{60}{30}

\usepackage{hyperref}
\DeclareRobustCommand{\new}[1]{\textcolor{blue}{#1}}

\definecolor{tensorblue}{RGB}{100, 149, 210}
\definecolor{tensorred} {RGB}{196,  80,  78}

\pgfmathsetmacro{\tnsc}{0.5}
\pgfmathsetmacro{\Step}   {\tnsc * 1.30}
\pgfmathsetmacro{\LegLen} {0.72 * \Step}
\pgfmathsetmacro{\Boxsz}  {\tnsc * 0.83}
\pgfmathsetmacro{\Bw}     {\tnsc * 2.20}
\pgfmathsetmacro{\Lw}     {\tnsc * 1.80}
\pgfmathsetmacro{\Nw}     {\tnsc * 1.00}
\pgfmathsetmacro{\Rc}     {\tnsc * 4.00}
\pgfmathsetmacro{\Nx}     {\tnsc * 0.17}
\pgfmathsetmacro{\Ny}     {\tnsc * 0.27}
\pgfmathsetmacro{\Na}     {\tnsc * 0.13}

\tikzset{
  tbox/.style = {draw, line width=\Bw pt, rounded corners=\Rc pt,
                 minimum size=\Boxsz cm, fill=#1},
  tbox/.default = tensorblue,
  leg/.style    = {line width=\Lw pt, line cap=round},
  lbl/.style    = {font=\large\itshape},
}

\newcommand{\Legs}[2]{%
  \draw[leg] (#1,#2) -- ++( \LegLen, \LegLen);%
  \draw[leg] (#1,#2) -- ++(-\LegLen, \LegLen);%
  \draw[leg] (#1,#2) -- ++( \LegLen,-\LegLen);%
  \draw[leg] (#1,#2) -- ++(-\LegLen,-\LegLen);%
}

\newcommand{\GateBox}[3]{%
  \node[tbox=#3] at (#1,#2) {};%
  \pgfmathsetmacro{\tmpNx}{#1+\Nx}%
  \pgfmathsetmacro{\tmpNy}{#2+\Ny}%
  \draw[line width=\Nw pt, line cap=round]%
    (\tmpNx,\tmpNy) -- ++(\Na,0) -- ++(0,-\Na);%
}

\newcommand{\LegLabels}[6]{%
  \draw[leg] (#1,#2) -- ++(-\LegLen, \LegLen) node[lbl,above left]  {$#3$};%
  \draw[leg] (#1,#2) -- ++( \LegLen, \LegLen) node[lbl,above right] {$#4$};%
  \draw[leg] (#1,#2) -- ++(-\LegLen,-\LegLen) node[lbl,below left]  {$#5$};%
  \draw[leg] (#1,#2) -- ++( \LegLen,-\LegLen) node[lbl,below right] {$#6$};%
}

\newcommand{\sect}[1]{\emph{#1}---}
\newcommand{\unilu}{Department of Physics and Materials Science, University of Luxembourg, L-1511 Luxembourg, Luxembourg}
\newcommand{\Uop}{U_\lambda}
\newcommand{\Usup}{\mathcal{U}_\lambda}   
\newcommand{\Iden}{\mathcal{I}}
\newcommand{\seed}{-i(\partial_\lambda \Uop)\Uop^\dagger}
\newcommand{\id}{\mathbb{I}}              

\begin{document}
\title{Counterdiabatic Quantum Circuits}

\author{Ankit W. Shrestha \orcidlink{0009-0008-5367-7062}}
    \email{ankit.wenjushrestha@uni.lu}
    \affiliation{\unilu}

\author{Budhaditya Bhattacharjee \orcidlink{0000-0003-1982-1346}}
    \email{budhaditya.bhattacharjee@uni.lu}
    \affiliation{\unilu}

\author{Adolfo~del Campo \orcidlink{0000-0003-2219-2851}}
    \email{adolfo.delcampo@uni.lu}
    \affiliation{\unilu}
    \affiliation{Donostia International Physics Center, E-20018 San Sebasti\'an, Spain}

\date{\today}

\begin{abstract}
Counterdiabatic driving provides a universal route to fast quantum control, but its standard formulation presupposes a Hamiltonian generator. We develop a Hamiltonian-free theory of the adiabatic gauge potential (AGP) for parameter-dependent unitary operators $U_\lambda$. Exact and regularized versions of the AGP follow from a Hermitian, banded linear system at a cost comparable to the Lanczos construction in the Hermitian case, requiring neither diagonalization nor a matrix logarithm. We demonstrate rapid convergence of the truncated Krylov approximation for the kicked top and for an integrable brickwork circuit built from unitary $\check{R}$ matrices, a setting in which no local effective Hamiltonian is available. Our results extend shortcuts to adiabaticity, and AGP-based diagnostics of chaos, to Floquet systems and quantum circuits.
\end{abstract}

\maketitle

Counterdiabatic driving (CD) constitutes a universal framework for fast quantum control, allowing one to speed up a physical process in quantum systems \cite{Demirplak&Rice_2003,Demirplak&Rice_2005,Demirplak&Rice_2008,Berry_2009}. In isolated systems described by a Hamiltonian, this is achieved by implementing an auxiliary counterdiabatic term, also known as the adiabatic gauge potential (AGP) \cite{GueryOdelin2019,Hatomura2024}. Shortcuts to adiabaticity by CD have been implemented in a wide variety of platforms, both at the single- and many-particle level. CD has also been harnessed for quantum computing, giving rise to a family of heuristic counterdiabatic quantum optimization algorithms \cite{Hegade2021,Chandarana2022,Hegade2022,Chandarana2024,Chandarana2026}.

It is by now widely appreciated that implementing the counterdiabatic term is costly, as the AGP generally involves nonlocal multiple-body interactions \cite{delCampo2012,Takahashi2013,Damski2014,Saberi2014}. A natural step to tackle such complexity has relied on approximating the AGP, by approaches ranging from variational methods \cite{Takahashi2013,Saberi2014,Takahashi2015,Sels2017} to Krylov subspace methods \cite{Claeys_2019,Takahashi_2024STA,Bhattacharjee_2023,Morawetz2025,Grabarits2025}. This has greatly broadened the applicability of CD, making possible implementations in digitized quantum computing \cite{Hegade2021,Chandarana2023,Chandarana2024,Visuri2026}, and recent progress has focused on constructing the AGP directly in a digitized way, without relying on auxiliary interactions \cite{Hatomura2026}.

Many settings of current interest are, however, specified directly by a unitary rather than by a Hamiltonian: Floquet and kicked systems, in which only the stroboscopic propagator is physical and the effective Hamiltonian $i\log U_\lambda$ is nonlocal, and unitary quantum circuits, whose natural building blocks are gates \cite{Bukov_FloquetEngineering,Claeys_UnitaryCircuits}. Faced with this scenario, the Hamiltonian formulation is at loggerheads with quantum circuit implementations.
This rationale motivates the quest for engineering CD evolution in the gate model of quantum computing, without ever introducing a Hamiltonian.

In this Letter, we provide a general framework for calculating CD in unitary systems without resorting to an effective Hamiltonian description. We derive a series solution for the AGP that uses the unitary and its derivative alone, and use Krylov space to find an effective truncation, in the spirit of \cite{Takahashi_2024STA,Bhattacharjee_2023, Shrestha_nonHermitian}.

\sect{Formulation.}
Our aim is to calculate the AGP that captures the diabatic transitions between the eigenstates of a unitary. These satisfy the eigenvalue relation
\begin{align}
    \Uop\ket{n(\lambda)} = e^{i\theta_n(\lambda)}\ket{n(\lambda)}\,,  \label{Eigval_relation}
\end{align}
where the unitary is a function of a tuning parameter $\lambda$, and the eigenphases $\theta_n(\lambda)$ are assumed non-degenerate. In the usual Hamiltonian formulation, the AGP is the generator of adiabatic deformations of the eigenbasis \cite{Demirplak&Rice_2003,Berry_2009}, $    A_\lambda = i\sum_{m\neq n}\ket{m}\braket{m}{\partial_\lambda n}\bra{n}$, and equivalently satisfies $[H_\lambda, i\partial_\lambda H_\lambda+[H_\lambda, A_\lambda]] = 0$ \cite{Kolodrubetz_2017}. Differentiating Eq.~\eqref{Eigval_relation} and projecting onto $\bra{m(\lambda)}$ gives
\begin{align}
    \bra{m(\lambda)}\partial_\lambda \Uop\ket{n(\lambda)} = &\left(e^{i\theta_n(\lambda)} - e^{i\theta_m(\lambda)}\right)\braket{m(\lambda)}{\partial_\lambda n(\lambda)} \nonumber \\
    &+ i(\partial_\lambda \theta_n)\, e^{i \theta_n(\lambda)}\,\delta_{mn}\,.
    \label{diff_eigval}
\end{align}
The last term encodes only the drift of the eigenphases and carries no information about transitions. In operator form, the remaining off-diagonal part reads
\begin{align}
    [\Uop, \,i\partial_\lambda \Uop + [\Uop, A_\lambda]] = 0\,.
    \label{AGP_comm_unitary}
\end{align}
For $m\neq n$, it is convenient to rewrite Eq.~\eqref{diff_eigval} as follows
\begin{align}
    \bra{m(\lambda)}A_\lambda \ket{n(\lambda)} = -i \frac{\bra{m(\lambda)}(\partial_\lambda \Uop)\Uop^\dagger\ket{n(\lambda)}}{e^{i(\theta_m - \theta_n)}-1}\,. \label{AGP_Floquet_matrix}
\end{align}
The above expression is exact. However, it requires diagonalization for each parameter value $\lambda$, which may not be feasible for larger systems. An alternative is to find a series expansion for $A_\lambda$, together with an effective truncation. Note that $(1-e^{i(\theta_m - \theta_n)})^{-1}$ is the sum of the geometric series $\sum_{r = 0}^\infty e^{i(\theta_m - \theta_n)r}$. Using this, we can denote $A_\lambda$ as the following infinite series
\begin{align}
    A_\lambda = -\lim_{s \to 1^-}\sum_{r = 0}^\infty s^{r}\, \Uop^r\left(\seed\right)(\Uop^\dagger)^r\,, \label{AGP_Floquet_infsum}
\end{align}
where $s$ is an Abel regulator added to ensure the convergence of the series (see Supplemental Material). It is straightforward to see that in the eigenbasis of $U_\lambda$, Eq.~\eqref{AGP_Floquet_infsum} reproduces Eq.~\eqref{AGP_Floquet_matrix}. Analogously to the Liouvillian superoperator that arises naturally in the Hamiltonian treatment, here we construct a superoperator $\Delta_U$ which serves the same purpose. For this, we begin by observing that the time-evolved operator in Eq.~\eqref{AGP_Floquet_infsum} is $-i(\partial_\lambda U)U^\dagger_\lambda$ instead of $\partial_\lambda U$. This suggests expressing Eq.~\eqref{AGP_comm_unitary} in terms of $i(\partial_\lambda U_\lambda)U^\dagger_\lambda$. If we define the following superoperator
\begin{align}
    \Delta_U (\cdot) = U (\cdot) U^\dagger - (\cdot)\,,\label{AGP_superoperator}
\end{align}
then the constraint relation in Eq.~\eqref{AGP_comm_unitary} can be written as
\begin{align}
    \Delta_U (i(\partial_\lambda U_\lambda)U^\dagger_\lambda) + \Delta^2_U(A_\lambda) = 0\,.\label{AGP_constraint}
\end{align}
The superoperator $\Delta_U$ is written in terms of the unitary superoperator $\mathcal{U}_\lambda(\cdot)= U_\lambda(\cdot)U_\lambda^\dagger$ as $\Delta_U = \mathcal{U}_\lambda - \mathcal{I}$. Note that the analogous constraint relation for the Hamiltonian AGP is $\mathcal{L}(i\partial_\lambda H) + \mathcal{L}^2(A_\lambda) = 0$, with the Liouvillian $\mathcal{L}(\cdot) = [H,\cdot]$, and that the same relation  holds for the unitary case with $\mathcal{L}_U(\cdot) = [U,\cdot]$, if one writes the AGP in terms of eigenvalue differences instead of eigenphase differences. The relation Eq.~\eqref{AGP_constraint} then implies that $A_\lambda$ can be written as $    A_\lambda = \Delta^{-1}_U\left(-i(\partial_\lambda U_\lambda)U^\dagger_\lambda\right)\,,$ provided this is restricted to the complement of the kernel of $\Delta_U$. 

The binomial theorem yields the exact analogue of the Hadamard (Baker--Campbell--Hausdorff) lemma,
\begin{align}
    \Uop^r\left(\seed\right)(\Uop^\dagger)^r = \sum_{k=0}^r \binom{r}{k}\, \Delta_U^k\!\left(\seed\right)\,, \label{AGP_Floquet_binom}
\end{align}
where $\Delta_U^k$ denotes the repeated action of $\Delta_U$. This motivates the following series form of the AGP operator\new{,}
\begin{align}
    A_\lambda = \sum_k c_k\,\Delta_U^k\!\left(\seed\right)\,.
    \label{AGP_series}
\end{align}
The coefficients $c_k$ can be fixed variationally \cite{Claeys_2019,Schindler_CDFloquet} or in Krylov space, as shown below. Table~\ref{table_key_expressions} collects these results alongside their static counterparts.

\sect{Krylov space method.}
As shown in \cite{Takahashi_2024STA,Bhattacharjee_2023}, an effective basis in which to write the terms of the series expansion is the Krylov basis, generated by the repeated action of the superoperator $\Delta_U$ followed by an appropriate orthonormalization algorithm. The algorithm requires neither diagonalization nor the logarithm of the unitary; it needs only the seed operator $\seed$, making it suitable for large many-body systems.

For the Hamiltonian formulation of the AGP, where the series consists of nested commutators with a Hermitian operator $H$, the Lanczos algorithm is an excellent tool for orthogonalizing the basis \cite{viswanath1994recursion}: it is iterative and requires orthogonalization against only two previous vectors at each step. It fails, however, when the generator of new basis vectors is non-Hermitian. For general generators, including non-Hermitian Hamiltonians and unitaries, bi-Lanczos or Arnoldi iteration generate a bi-orthonormal or orthonormal basis, at additional cost \cite{Bhattacharya2022arnoldi,Bhattacharjee2023dissSYK,Nizami2023Floquet,Bhattacharya2023biLanczos,Zhou2025manybodychaos}. Exploiting instead the unitarity of the generator, Ref.~\cite{Trunin2025_KrylovFloquet} employed the Szeg\H{o} algorithm, whose cost is exactly that of Lanczos, while bringing unitaries to the upper-Hessenberg form, which is otherwise obtained only through the more expensive Arnoldi iteration. In our formalism, the generator $\Delta_U$ is not unitary but is related to the unitary superoperator by $\Delta_U = \Usup - \Iden$. Since the two differ by a multiple of the identity, they generate the same Krylov subspace from a given seed, and we may build the basis from $\Usup$. The Szeg\H{o} algorithm can be further optimized to produce, at the same computational cost, a pentadiagonal representation of $\Usup$ \cite{Trunin2025_KrylovFloquet}. The resulting basis is the Cantero--Moral--Vel\'azquez (CMV) basis \cite{Cantero2003} (see Supplemental Material for details of the algorithm).

The superoperator $\Delta_U$ is normal, with $\Delta_U^\dagger\Delta_U = 2\Iden - \Usup - \Usup^\dagger$. The superoperator $\Delta_U^\dagger\Delta_U$ is Hermitian, positive semidefinite and inherits the pentadiagonal structure of $\Usup$. The AGP then follows from the normal equations associated with Eq.~\eqref{AGP_constraint},
\begin{align}
    \Delta_U^\dagger\,\Delta_U\,|A_\lambda) = -i\,\Delta_U^\dagger\,\big|(\partial_\lambda \Uop)\Uop^\dagger\big)\,,
    \label{AGP_normal_eq}
\end{align}
a banded Hermitian system of dimension $K$ whose minimum-norm solution reproduces Eq.~\eqref{AGP_Floquet_matrix} exactly, with $\Delta_U^\dagger$ on the right-hand side annihilating the eigenphase-drift component of the seed.

\sect{Regularized AGP.}
We have thus far discussed the exact AGP, which is what transitionless driving requires. However, recent studies have used the AGP (and in particular its norm) as a diagnostic tool. The AGP norm captures quantum phase transitions in many-body systems such as the transverse-field Ising chain, being proportional to the fidelity susceptibility \cite{delCampo2012,Funo2017,Bhattacharjee_2023}. The growth of the rescaled AGP norm $\|A_\lambda\|^2/L$ with the system size $L$ is also a sensitive probe of quantum chaos \cite{Pandey_ChaosDiagnostic}. Similarly,  in Floquet systems,  the AGP detects the driving frequency above which the system enters a robust prethermal phase with exponentially long heating times \cite{Bhattacharjee_FloquetHeating}.

For such applications, it is crucial to `regularize' the AGP, so as to avoid large contributions from near-degenerate terms (see the denominator of Eq.~\eqref{AGP_Floquet_matrix}). This is particularly important for Floquet systems, since their eigenphases lie on a compact unit circle and near degeneracies therefore occur more frequently. Such divergent terms are more likely to appear in larger systems, and in integrable systems, where there is no level repulsion among the eigenstates. We can write the regularized AGP as
\begin{align}
    \Big[A_\lambda^{(R)}\Big]_{mn} = -i\; \frac{\left(e^{-i(\theta_m - \theta_n)}-1\right)\; \Big[(\partial_\lambda \Uop)\Uop^\dagger\Big]_{mn}}{\big|e^{i(\theta_m - \theta_n)}-1\big|^2 + \eta^2} \,, \label{regAGP_matrix}
\end{align}
where $\eta$ is the regulator. It reduces to the exact AGP in the limit $\eta \to 0$, and the numerator makes the diagonal $m=n$ elements vanish identically, so that Eq.~\eqref{regAGP_matrix} may be used for all $m,n$. To approximate the regularized AGP we simply add a Tikhonov term to Eq.~\eqref{AGP_normal_eq} and solve in Krylov space (see Supplemental Material),
\begin{align}
    \left(\Delta_U^\dagger \Delta_U + \eta^2\Iden\right)\big|A_\lambda^{(R)}\big) = -i\, \Delta_U^\dagger \,\big|(\partial_\lambda \Uop)\Uop^\dagger\big)\,. \label{regAGP_eqn}
\end{align}
The left-hand side is pentadiagonal and strictly positive definite for $\eta>0$, so Eq.~\eqref{regAGP_eqn} is solved in $\mathcal{O}(K)$ operations. Truncating the CMV basis at $M<K$ vectors provides a natural truncation for the series approximation.

This regularization, which is in the spirit of the regulator employed for Hamiltonian AGP, is different from the Abel regulator in Eq.~\eqref{AGP_Floquet_infsum}. This regulated AGP can also be interpreted as a Floquet infinite sum with forward-backward evolutions
\begin{align}
    A^{(R)}_\lambda =& - \frac{\sigma}{1+\sigma}\sum_{r \in \mathbb{Z}}\text{sgn}\left(r + \frac{1}{2}\right)\sigma^{\left\vert r + \frac{1}{2}\right\vert - \frac{1}{2}}\notag\\ &\times\mathcal{U}^r_\lambda \left(\seed\right)\,,
\end{align}
where we have the backward evolution $\mathcal{U}^{-r}_\lambda(X) = \left(\Uop^\dagger\right)^r X \Uop^r$, the forward evolution $\mathcal{U}^{r}_\lambda(X) = \Uop^r X \left(\Uop^\dagger\right)^r$ and the regulator $\eta^2 = (1-\sigma)^2/\sigma$ such that $A_\lambda = \lim_{\sigma \to 1^{-}} A_\lambda^{(R)}$. The range of $r$ is over all integers, including $r=0$. This is the natural counterpart of the Hamiltonian AGP expression~\cite{Kolodrubetz_2017,Claeys_2019,Pandey_ChaosDiagnostic}. In what follows, we study how this Krylov space approximation converges for the regularized AGP in Floquet systems.

\sect{Application to the Kicked Top.}
As an illustrative example, we study the quantum kicked top, which describes a top undergoing constant precession around a magnetic field and is subject to periodic delta-function kicks  \cite{Haake1987}. It is a paradigmatic model of quantum chaos which has a well-defined classical limit and shows integrable to chaotic transition with a tunable parameter \cite{Haake1987},  demonstrated experimentally in \cite{Chaudhury2009}. The Hamiltonian for the system can be written as
\begin{align}
    H(t) = \frac{p}{\tau}J_y + \frac{k}{2j}J_z^2 \sum_{n \in \mathbb{Z}}\delta(t - n\tau)\,,
\end{align}
where $J_y, J_z$ are the spin-$j$ angular momentum operators and $\tau$ defines the kick period, that we set to 1. The Floquet unitary that describes the time-evolution under this Hamiltonian is given by $U(k,p) = e^{-i(k/2j)J_z^2}e^{-ipJ_y}$.
The model is integrable for small values of $k$ and becomes chaotic as we tune it to larger values. We will set the precession angle $p=0.9$ for the calculations. Treating $k$ as the tuning parameter $\lambda$, the seed operator for the AGP can be written in the closed form as $G = (-1/2j)J_z^2$.
\begin{figure*}[ht]
    \centering
    \includegraphics[width=0.32\linewidth]{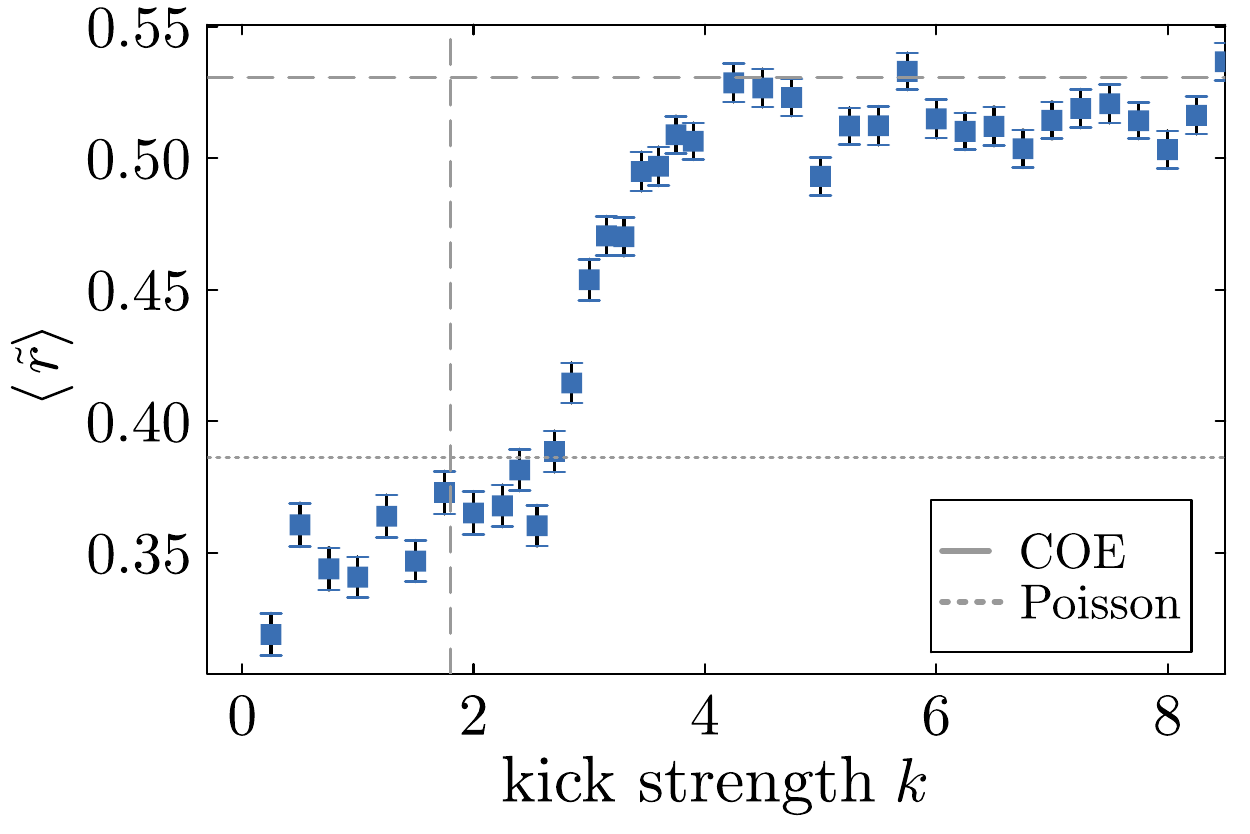}
        \includegraphics[width=0.32\linewidth]{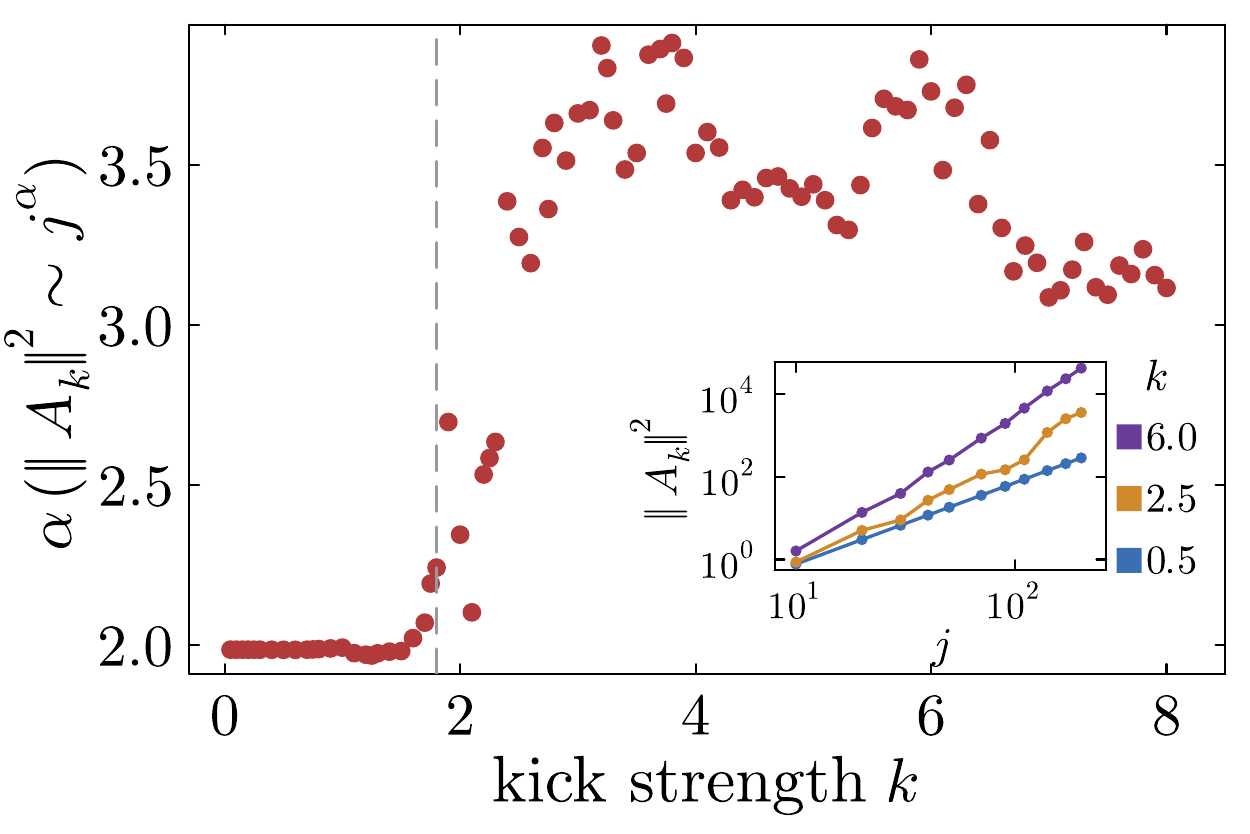}
    \includegraphics[width=0.32\linewidth]{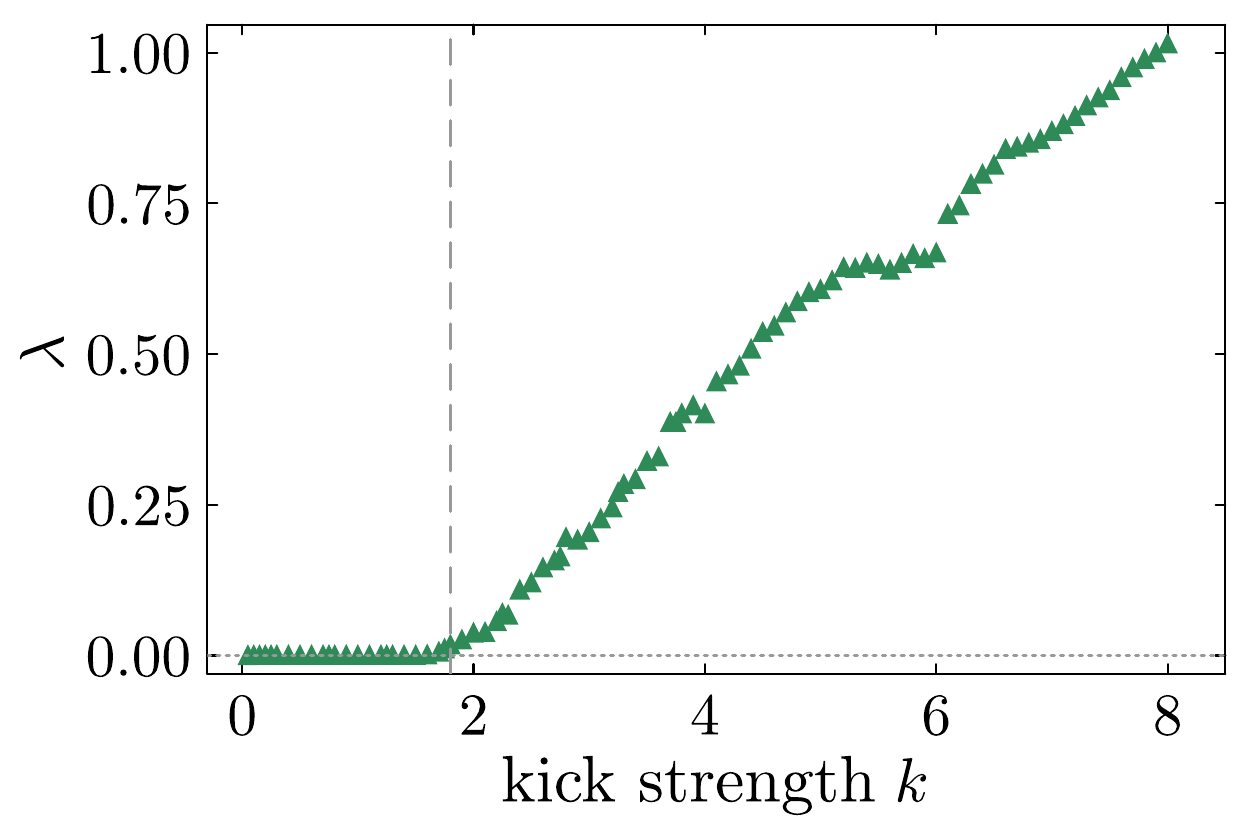}
    \caption{Mean level spacing ratio (left), AGP norm scaling exponent with system size (middle), and Lyapunov exponent (right) capturing the integrable to chaos transition in quantum and classical regimes of the kicked top. Vertical lines mark the onset of classical chaos and the inset (middle) shows how AGP norm scales with the system size.}
    \label{fig:1}
\end{figure*}

We study two things, scaling of AGP norm with the system size as a sensitive probe of chaos (in line with \cite{Pandey_ChaosDiagnostic}) and convergence of approximate regularized AGP towards the true value demonstrating the efficiency of the series approximation. The integrability-to-chaos transition in the quantum regime can be captured by the mean level spacing ratio  $\langle \tilde r\rangle = \frac{1}{D} \sum_{n=1}^D \frac{\min(s_n, s_{n-1})}{\max(s_n, s_{n-1})}$,
where $s_n = (\varphi_n - \varphi_{n-1})$ is the spacing between consecutive eigenphases of the unitary folded to $[0, 2\pi)$. Similarly, we can define the classical limit of the model with $j\to \infty$ and capture the same transition using the Lyapunov exponent in the phase space~\cite{Haake1987}. In \cite{Pandey_ChaosDiagnostic}, it was demonstrated that scaling the regularized AGP norm with the system size ($||A^{(R)}||^2 \sim j^\alpha$) is a sensitive probe of quantum chaos. Note that the regularization parameter should scale as $\eta(j) \sim \pi/j$ to keep it comparable to the mean level spacing. Figure \ref{fig:1} shows the transition in both quantum and classical regimes while tuning the parameter $k$ using the three different measures. We observe that the exponent $\alpha$ captures the transition to chaos earlier than the mean spacing ratio in the quantum regime.

\begin{figure}[t]
 \centering
    \includegraphics[width=0.49\linewidth]{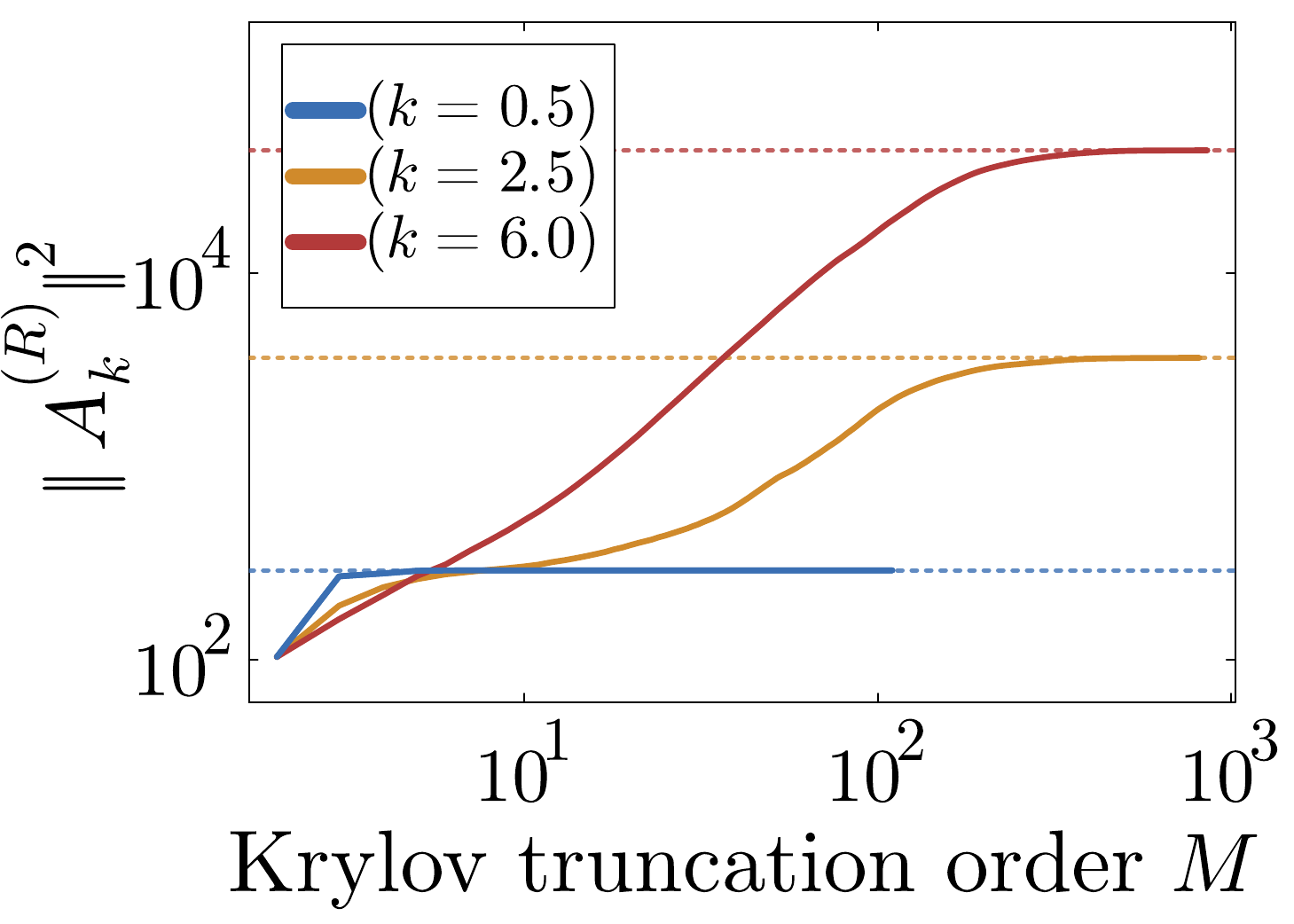}
\includegraphics[width=0.49\linewidth]{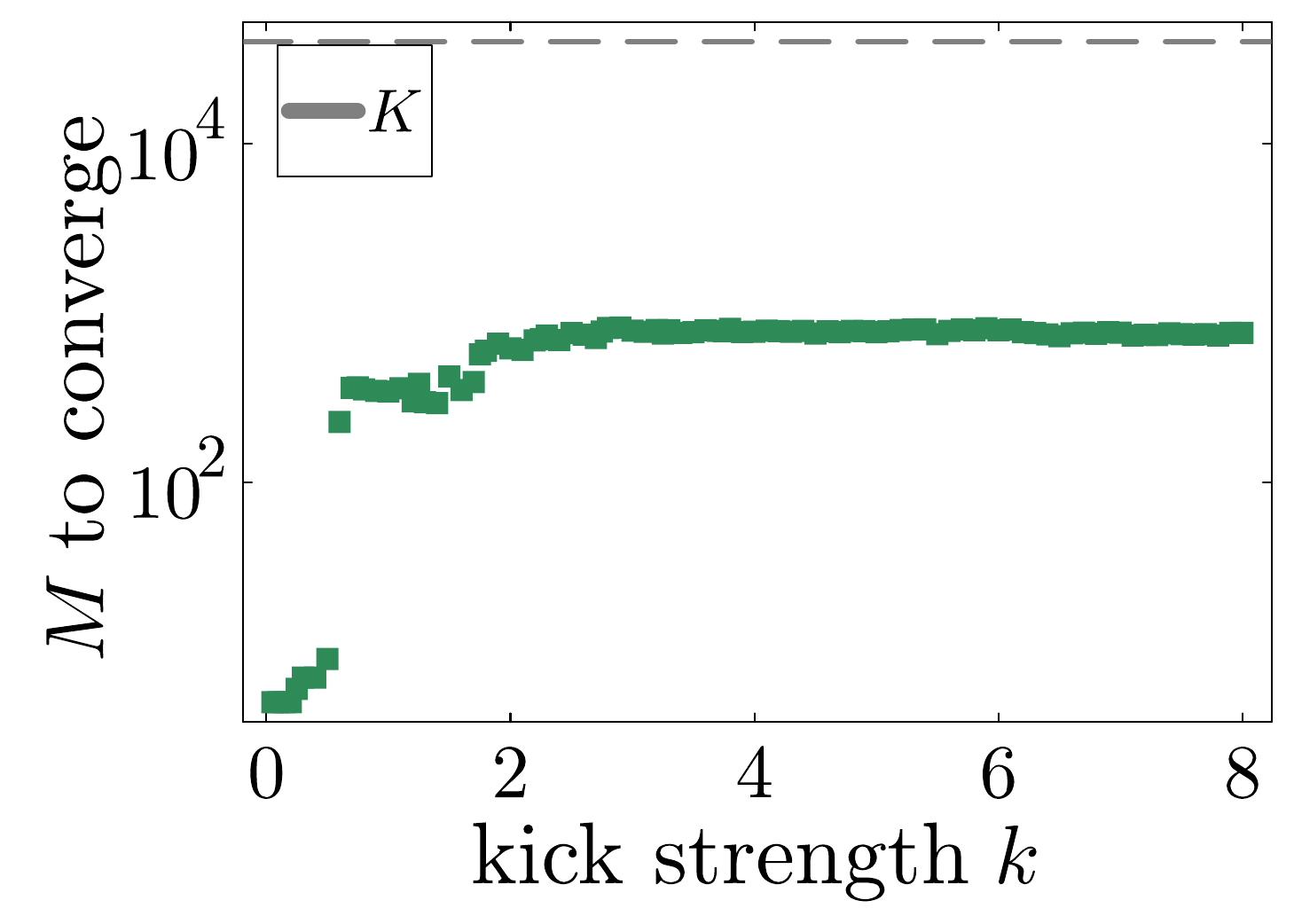}
    \caption{Monotonic convergence of AGP norm towards the true value as more terms are added in the series expansion (left). Note that the AGP norm converges to the true value (with relative error $\lesssim 10^{-4}$) much earlier than the full Krylov space dimension (right).}
    \label{fig:2}
\end{figure}

In Fig. \ref{fig:2}, we study how the approximate AGP norm obtained from the finite truncation of the Krylov basis vectors approaches the true AGP norm derived from exact diagonalization. We note that the convergence towards the true value is monotonic as more terms are added in the series expansion and the convergence to the true value (up to the relative error $\lesssim 10^{-4}$) happens much earlier than the full Krylov space dimension.

\sect{Application to a brickwork circuit.} 
Unitary circuits serve as a minimal model for many-body dynamics in which local interactions dominate. Such interactions are modeled by unitary gates acting on $k$-local sites, and the dynamics is generated through the repeated action of a chain of such gates. They also arise naturally in the context of Hamiltonian simulation on quantum computers, where the time-evolution operator is discretized into unitary gates after Trotterization~\cite{Lloyd_QuantumSimulator}. We consider one particular case, the `brickwork circuit', in which the two-site unitary gates $U_{j,j+1}$ are arranged such that
\begin{align}
    &U = U_{\text{odd}}U_{\text{even}}\,,  \nonumber\\
    &U_{\text{odd}} = \prod_{j} U_{2j-1, 2j},\;\;
    U_{\text{even}} = \prod_{j} U_{2j, 2j+1}\,. \label{brickwork_circuit}
\end{align}
In graphical notation, a single gate (left) and the unitary circuit it builds (right) read
\begin{center}
\resizebox{\columnwidth}{!}{%
\begin{tikzpicture}
  \pgfmathsetmacro{\MidY}{1.5*\Step}
  %
  \node[font=\large, anchor=east] at (-\tnsc*1.55, \MidY) {$U_{ab,cd}\;=$};
  \LegLabels{0}{\MidY}{a}{b}{c}{d}
  \GateBox{0}{\MidY}{tensorblue};
  %
  \begin{scope}[xshift=3.05cm]
    \foreach \r in {1,3} {
      \pgfmathsetmacro{\gy}{\r*\Step}
      \foreach \c in {0,...,3} {
        \pgfmathsetmacro{\gx}{\c*2*\Step}
        \Legs{\gx}{\gy}
      }
    }
    \foreach \r in {0,2} {
      \pgfmathsetmacro{\gy}{\r*\Step}
      \foreach \c in {0,...,3} {
        \pgfmathsetmacro{\gx}{(2*\c+1)*\Step}
        \Legs{\gx}{\gy}
      }
    }
    \foreach \r in {1,3} {
      \pgfmathsetmacro{\gy}{\r*\Step}
      \foreach \c in {0,...,3} {
        \pgfmathsetmacro{\gx}{\c*2*\Step}
        \GateBox{\gx}{\gy}{tensorblue}
      }
    }
    \foreach \r in {0,2} {
      \pgfmathsetmacro{\gy}{\r*\Step}
      \foreach \c in {0,...,3} {
        \pgfmathsetmacro{\gx}{(2*\c+1)*\Step}
        \GateBox{\gx}{\gy}{tensorblue}
      }
    }
    \node[font=\large, anchor=east] at (-\tnsc*1.40, \MidY) {$U(t)\;=$};
  \end{scope}
\end{tikzpicture}}
\end{center}
Each leg acts on a local two-dimensional Hilbert space, and the circuit shown represents two time steps for the $L=8$ lattice sites. The Krylov space dynamics of such models was studied in \cite{Suschland_KrylovTrotter}. An example of this structure arises in the integrable Trotterization of Heisenberg spin chains, where each gate is represented by an $\check{R}$ matrix satisfying the braided Yang--Baxter equation \cite{Claeys_UnitaryCircuits,Vanicat_IntegrableTrotterization}. The formulation presented above is particularly well suited to calculating the AGP in such models, since it requires only the unitary operator and bypasses the need for an effective Hamiltonian description, which is not natural in this context.

Consider a spin-$1/2$ anisotropic Heisenberg (XXZ) chain,
\begin{align}
    H_{XXZ} = J \sum_{j=1}^{L-1} \left(\sigma_j^x\sigma_{j+1}^x + \sigma_j^y\sigma_{j+1}^y + \Delta \,\sigma_j^z\sigma_{j+1}^z\right)\,,
\end{align}
where $\Delta$ controls the anisotropy. The unitary dynamics generated by this Hamiltonian can be Trotterized into the brickwork circuit, where each gate is represented by the matrix
\begin{align}
\check{R}(\theta,\gamma) = 1 \oplus \frac{1}{\nu}
\begin{pmatrix}
\sin\gamma & i\sinh\theta \\[2pt]
i\sinh\theta & \sin\gamma
\end{pmatrix} \oplus 1 \,,
\label{R_matrix_unitary}
\end{align}
with $\nu(\theta,\gamma) = \sin(\gamma+i\theta) = \sin{\gamma} \cosh{\theta} + i \cos{\gamma}\sinh{\theta}$ as the normalization factor (see Supplemental Material). Since $|\nu|^2=\sin^2\gamma+\sinh^2\theta$, the inner $2\times2$ block is unitary and $\check{R}^\dagger\check{R}=\id_4$. The tunable parameter is $\theta \in \mathbb{R}$. As $\theta \to 0$, each gate reduces to the identity operator, and the circuit acts trivially on the initial state. For small values of $\theta$, the circuit acts as a Hamiltonian simulator for the XXZ Hamiltonian. Similarly, each gate approaches a SWAP-like structure for large values of $\theta$. The Hermitian seed operator $-i(\partial_\theta U)U^\dagger$ is built from
local operators as
\begin{equation}
  -i(\partial_\theta U)U^\dagger
  = G_\theta^{\mathrm{odd}}
  + U_{\mathrm{odd}}\, G_\theta^{\mathrm{even}}\, U_{\mathrm{odd}}^\dagger ,
  \label{eq:seed}
\end{equation}
where $G_\theta^{\mathrm{even/odd}} = \sum_{j\in\mathrm{even/odd}} g_{\theta,j}$.
Here, $g_{\theta,j}$ denotes the two-site operator
$g_\theta^{\mathrm{loc}}(\gamma) \equiv -i(\partial_\theta \check{R})\check{R}^\dagger$
acting on sites $j$ and $j+1$, tensored with the identity on all other sites.
\begin{figure}[t]
    \centering
    \includegraphics[width=0.49\linewidth]{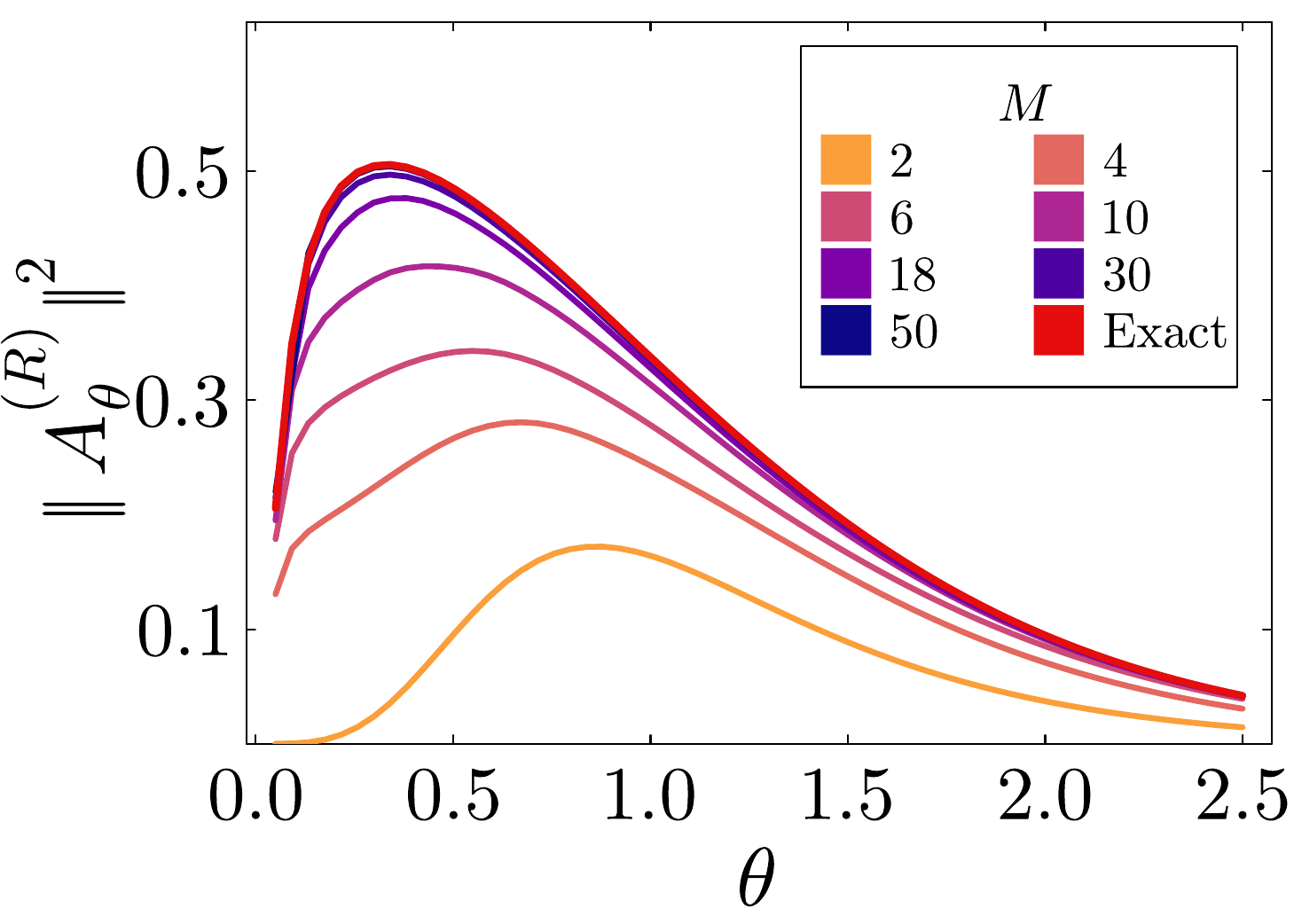}
    \includegraphics[width=0.49\linewidth]{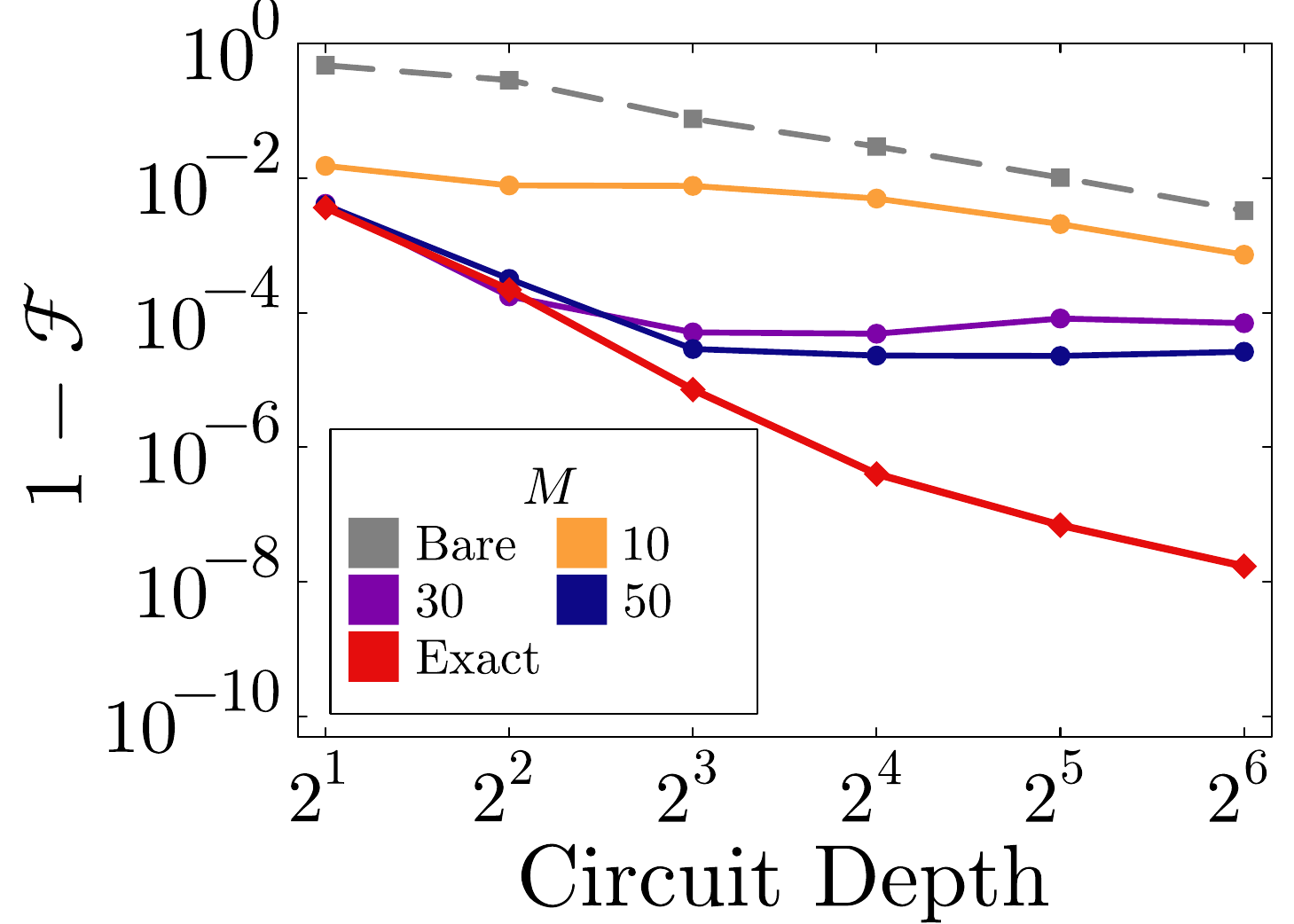}
   \caption{(Left) Regularized AGP norm for the brickwork circuit with $\check{R}$ matrices as unitary gates ($L=12$, $\gamma=1.0$), for several truncation orders $M$ of the Krylov space. (Right) Infidelity of a state driven from $\theta = 1.5 \to 0.2$ across different circuit depths with and without the counterdiabatic gates.}
    \label{fig: AGP_norm_brickwork}
\end{figure}
Figure~\ref{fig: AGP_norm_brickwork} (left) shows the regularized AGP norm as a function of $\theta$. The anisotropy parameter is fixed at $\gamma = 1.0$, i.e.,\ $\Delta = \cos\gamma \approx 0.54$. Unitarity of $\check{R}$ requires $\gamma\in\mathbb{R}$ and therefore $|\Delta|<1$. As a result, the integrable Trotterization is confined to the gapless (critical) regime of the XXZ chain. The AGP norm captures the size of the operator required to mitigate diabatic transitions between the eigenstates. Here, the gates are close to the identity for small values of $\theta$, and they barely change with $\theta$ once $\theta$ is large compared to $\gamma$. The interesting dynamics lies at intermediate values, where the circuit acts as a quantum simulator for the XXZ Hamiltonian, and this is nicely captured by the AGP norm. We can observe how different truncation orders in the series expansion (equivalently, of the Krylov space) approach the true value of the norm. In Fig.~\ref{fig: AGP_norm_brickwork} (right), we show how a state can be adiabatically driven across the parameter values $\theta = 1.5 \to 0.2$ using counterdiabatic gates. We measure the infidelity $1 - \mathcal{F} = 1 - \left|\bra{\psi_g(\theta_{\min})}\,\prod_{i=1}^{S} U(\theta_i)e^{-i\delta\theta\,A(\theta_i)}\,\ket{\psi_g(\theta_{\max})}\right|^2$, where $S$ is the circuit depth. The eigenstates were ordered unambiguously using the Kato Hamiltonian as prescribed in \cite{Schindler_GeometricFloquet}. Compared to bare evolution, counterdiabatic gates suppress infidelity by up to six orders of magnitude.  We also show how approximate AGP with finite truncation, which involves relatively local gates for easy implementation, can achieve the suppression efficiently.

\sect{Conclusion.}
Counterdiabatic driving has so far been tied to a Hamiltonian, whereas Floquet systems and quantum circuits rely on unitary gates. We have shown that the adiabatic gauge potential, the control field that suppresses unwanted transitions as a parameter is varied, follows directly from a unitary and its derivative, as a series generated by repeated action of the unitary. Truncating after a handful of terms already recovers the exact result, as a kicked top and an integrable circuit demonstrate. Unitarity of the generator makes its Krylov representation banded, so the cost matches that of the Hamiltonian approach. This opens counterdiabatic control, and the use of the gauge potential as a probe of chaos and criticality, to
 Floquet engineering, digital quantum simulation, and protocols compiled directly into gates.

{\it Acknowledgements.---}
This project was supported by the Luxembourg National Research Fund (FNR Grant No
C24/MS/18940482/STAOpen).

{\it Data Availability.---} Details of the numerical calculations with codes are available on GitHub~\cite{Wenju_CDCircuits}.
\bibliography{STA_Refs}

\clearpage
\onecolumngrid 
\section*{End Matter}
\subsection{Comparison of Hamiltonian and Unitary AGP}

\begin{table}[H]
\centering
\setlength{\tabcolsep}{16pt}
\renewcommand{\arraystretch}{1.8}
\resizebox{\textwidth}{!}{%
\begin{tabular}{lll}
\hline \hline
                            & Hamiltonian $H_\lambda$ & Unitary $U_\lambda$ \\ \hline
AGP  commutation relation               &   $[H_\lambda, i\partial_\lambda H_\lambda+[H_\lambda, A_\lambda]] = 0$                      &   $[U_\lambda, i\partial_\lambda U_\lambda + [U_\lambda, A_\lambda]] = 0$                  \\
   &or, sufficiently, $\partial_\lambda H_\lambda = i[H_\lambda, A_\lambda]$& or, sufficiently, $\partial_\lambda U_\lambda = i[U_\lambda, A_\lambda]$ \\ \hline
Solution                    &    $A_\lambda = \mathcal{L}^{+}(-i\partial_\lambda H_\lambda)$                     &    $A_\lambda = \Delta_U^{+}(-i(\partial_\lambda U_\lambda)U_\lambda^\dagger)$                 \\
&with $\mathcal{L}(\cdot) = [H_\lambda, \cdot]$& with $\Delta_U(\cdot) = U_\lambda(\cdot) U_\lambda^\dagger -  (\cdot) $\\ \hline
Exact (regularized) expansion    &  $A_\lambda = -\frac{1}{2} \lim_{\mu \to 0^{+}} \int_{-\infty}^\infty {\rm sgn}(t)\,e^{-\mu |t|}\,$                       &     $A_\lambda = -\lim_{\sigma \to 1^{-}} \frac{\sigma}{1+\sigma}\sum_{r \in \mathbb{Z}}\text{sgn}\left(r + \frac{1}{2}\right)\sigma^{\left\vert r + \frac{1}{2}\right\vert - \frac{1}{2}}$                   \\
& \hspace{20pt} $\times \;e^{iH_\lambda t} (\partial_\lambda H_\lambda)\,e^{-iH_\lambda t}\, dt$&  \hspace{20pt}$\times\;\mathcal{U}^r_\lambda \left(\seed\right)$ \\ \hline
Approximate series solution &   $A_\lambda = i\sum_k c_k (\lambda)\; \mathcal{L}_\lambda^{2k-1}(\partial_\lambda H_\lambda)$   &  $ A_\lambda = \sum_k c_k(\lambda) \,\Delta_U^k(-i(\partial_\lambda U_\lambda)U_\lambda^\dagger)$  \\
\hline
Krylov generator & $\mathcal{L}$ Hermitian $\Rightarrow$ Lanczos (tridiagonal) & $\mathcal{U}_\lambda$ unitary $\Rightarrow$ Szeg\H{o}/CMV (pentadiagonal)\\
\hline \hline
\end{tabular}%
}
\caption{Summary of key expressions for the AGP calculation in Hamiltonian $H_\lambda$ and unitary $U_\lambda$ systems. Here $\mathcal{L}(\cdot)=[H_\lambda,\cdot]$ and $\Delta_U=\mathcal{U}_\lambda-\mathcal{I}$ are the respective generators. The Moore--Penrose inverse of $X$ is denoted by ${X}^{+}$, and projects out the kernel (diagonal sector) of each generator. $\mu\to0^+$ in the Hamiltonian column and $\sigma\to1^-$ in the unitary column implement the same regularization.}
\label{table_key_expressions}
\end{table}

\newpage\clearpage

\setcounter{equation}{0}
\renewcommand{\theequation}{S\arabic{equation}}

\begin{center}
{\large\bfseries Supplemental Material: Counterdiabatic Quantum Circuits\par}
\vspace{10pt}
Ankit W. Shrestha \orcidlink{0009-0008-5367-7062}$^{1}$,\; 
Budhaditya Bhattacharjee \orcidlink{0000-0003-1982-1346}$^{1}$,\; and 
Adolfo~del Campo \orcidlink{0000-0003-2219-2851}$^{1,2}$\\[3pt]
$^{1}$\textit{\unilu}\\
$^{2}$\textit{Donostia International Physics Center, E-20018 San Sebasti\'an, Spain}
\end{center}

\section{AGP for Floquet systems} \label{app_AGP_Floquet}
In what follows, we detail some mathematical steps for the calculations in the main text. First, we show that the infinite series expansion in Eq.~\eqref{AGP_Floquet_infsum} correctly reduces to the AGP in Eq.~\eqref{AGP_Floquet_matrix}. Starting with the matrix element, we get
\begin{align}
    \bra{m(\lambda)}A_\lambda \ket{n(\lambda)} &= -\lim_{s \to 1^-}\sum_{r = 0}^\infty s^{r}\, \bra{m(\lambda)}\Uop^r\left(\seed\right)(\Uop^\dagger)^r \ket{n(\lambda)} \nonumber \\
    &= -\lim_{s \to 1^-}\sum_{r = 0}^\infty s^{r}\,e^{i(\theta_m - \theta_n)r} \bra{m(\lambda)}\left(\seed\right)\ket{n(\lambda)}\,,
\end{align}
where we have used the eigenvalue relation of Eq.~\eqref{Eigval_relation} and $s$ is the Abel regulator. Focusing only on the off-diagonal terms $m \neq n$, we can evaluate the infinite sum as
\begin{align}
    \lim_{s \to 1^-}\sum_{r = 0}^\infty \left(s\, e^{i(\theta_m - \theta_n)}\right)^r &= \lim_{s \to 1^-} \frac{1}{1 - s\, e^{i(\theta_m - \theta_n)}} = \frac{1}{1 - e^{i(\theta_m - \theta_n)}}\,,
\end{align}
which is valid since $|s\,e^{i(\theta_m - \theta_n)}| < 1$. Plugging this back gives the correct expression
\begin{align}
    \bra{m(\lambda)}A_\lambda \ket{n(\lambda)} = -i \frac{\bra{m(\lambda)}(\partial_\lambda \Uop)\Uop^\dagger\ket{n(\lambda)}}{e^{i(\theta_m - \theta_n)}-1}\,.
\end{align}

Next, we will show the validity of relation in Eq.~\eqref{AGP_Floquet_binom}. Recall that the adjoint superoperator $\Usup$ acts as $\Usup(X) = \Uop X \Uop^\dagger$, and that it is related to $\Delta_U$ by $\Usup = \Delta_U + \Iden$, where $\Iden$ is the identity superoperator. The left hand side of the expression in Eq.~\eqref{AGP_Floquet_binom} is simply
\begin{align}
    \Usup^r(X) = (\Delta_U + \Iden)^r(X) = \sum_{k=0}^r \binom{r}{k}\,\Iden^{\,r-k}\Delta_U^k(X)\,,
\end{align}
using the binomial expansion. The symbol $\mathcal{L}$ is reserved throughout for the Hamiltonian commutator superoperator $\mathcal{L}(\cdot)=[H_\lambda,\cdot]$ of Table~\ref{table_key_expressions}.

We will finally show the validity of Eq.~\eqref{regAGP_eqn} to approximate the regularized AGP in the operator space. The matrix elements of $\Delta_U(X)$ in the eigenbasis of $U$ can be written as
\begin{align}
    \Big[\Delta_U(X)\Big]_{mn} = \left(e^{i(\theta_m - \theta_n)}-1\right)X_{mn}\,.
\end{align}
The superoperator $\Delta_U$ is not Hermitian, although it is normal since $\Usup$ is unitary. Its adjoint is given by
\begin{align}
    \Big[\Delta_U^\dagger( X)\Big]_{mn} = \left(e^{-i(\theta_m - \theta_n)}-1\right)X_{mn}\,.
\end{align}
Equivalently, $\Delta_U^\dagger\Delta_U = 2\Iden - \Usup - \Usup^\dagger$, which is Hermitian and positive semidefinite. Now, we can calculate
\begin{align}
    \Big[\left(\Delta_U^\dagger \Delta_U + \eta^2\Iden\right)\big(A_\lambda^{(R)}\big) \Big]_{mn} = \Big[ \Delta_U^\dagger \left(\seed\right) \Big]_{mn}\,,
\end{align}
which gives the exact expression in Eq.~\eqref{regAGP_matrix}
\begin{align}
    \Big[A_\lambda^{(R)}\Big]_{mn} = -i\; \frac{\left(e^{-i(\theta_m - \theta_n)}-1\right)\; \Big[(\partial_\lambda \Uop)\Uop^\dagger\Big]_{mn}}{\big|e^{i(\theta_m - \theta_n)}-1\big|^2 + \eta^2} \,,
\end{align}
with $\eta$ the regulator introduced in the main text.

\section{Szeg\H{o} Algorithm and AGP in CMV Basis}
In this section, we collect the details of the Szeg\H{o} recursion and of the resulting Cantero--Moral--Vel\'azquez (CMV) basis, in which the superoperator $\Usup$ is pentadiagonal (see \cite{Trunin2025_KrylovFloquet} for a pedagogical discussion). We start by mapping operators $O$ to vectors $|O)$ in the operator Hilbert space equipped with the inner product $(O_a|O_b) =  \Tr[O_a^\dagger O_b]/D$, where $D$ is the Hilbert space dimension. Thus, the unitary superoperator acts as $\Usup|P) = |\Uop P \Uop^\dagger)$. Starting from the normalized seed operator $|\Phi_0) = |\tilde{\Phi}_0) =  \mathcal{N}\,|\seed)$, with $\mathcal{N} = \big\|\,|\seed)\,\big\|^{-1}$, the Szeg\H{o} algorithm iteratively proceeds as follows:
\begin{align}
  \bar\alpha_n = (\tilde\Phi_n|&\,\mathcal{U}_\lambda\,|\Phi_n) ,
  \qquad
  \rho_n = \sqrt{1-|\alpha_n|^2} ,
   \nonumber\\
  \rho_n|\Phi_{n+1}) &= \mathcal{U}_\lambda|\Phi_n) - \bar\alpha_n|\tilde\Phi_n) ,
  \nonumber\\
  \rho_n|\tilde\Phi_{n+1}) &= |\tilde\Phi_n) - \alpha_n\,\mathcal{U}_\lambda|\Phi_n) ,
  \label{eq:em-szego}
\end{align}
where, $\alpha_n$ are the Verblunsky coefficients. Recursion stops automatically with $\rho_{K-1} = 0$ at the maximum dimension of the Krylov space $K$, or can be halted manually at $M<K$ for approximate AGP. The unitary superoperator can be written in terms of $\{\alpha_n, \rho_n\}$ in Krylov space in two different representations. In the Szeg\H{o} basis $\{ |\Phi_n)\}$, the superoperator takes an upper Hessenberg form
\begin{align}
  (\Phi_m|\,\mathcal{U}_\lambda\,|\Phi_n) =
  \begin{cases}
    -\,\bar\alpha_n\,\alpha_{m-1}\displaystyle\prod_{k=m}^{n-1}\rho_k , & m < n+1 , \\[6pt]
    \rho_n , & m = n+1 , \\[2pt]
    0 , & m > n+1 ,
  \end{cases}
  \label{eq:em-hess}
\end{align}
with the convention $\alpha_{-1} = -1$. We can reorder the basis as $|P_{2n}) = \mathcal{U}^{-n}_\lambda|\tilde\Phi_{2n}) ,
  \;
  |P_{2n+1}) = \mathcal{U}^{-n}_\lambda|\Phi_{2n+1})$
to get the Cantero-Moral-Velazquez (CMV) basis, where the superoperator has a more compact pentadiagonal form
\begin{align}
    (P_m| \mathcal{U_\lambda} |P_n) = \sum_{k=0}^{K-1} \Xi_{mk} \Lambda_{kn}\,.
\end{align}
Here, $\Xi = \Theta_0 \oplus \Theta_2 \oplus \Theta_4\oplus \dots$ and $\Lambda= 1\oplus \Theta_1 \oplus \Theta_3\oplus \dots$ are $K\times K$ matrices formed of $2\times 2$ dimensional blocks $\Theta_n$ defined as
\begin{equation}
  \Theta_n =
  \begin{pmatrix}
    \bar\alpha_n & \rho_n \\
    \rho_n & -\alpha_n
  \end{pmatrix} ,
  \qquad \det\Theta_n = -1 \,,
  \label{eq:em-theta}
\end{equation}
and closed with $1\times 1$ identity block at the end. The approximate AGP can be written in the CMV basis  as $ |A_\lambda) = \sum_{n=0}^{M-1} c_n|P_n)$ using the first $M$ basis vectors, and the coefficients $c_n$ can be fixed by solving Eq.~\eqref{AGP_normal_eq} or Eq.~\eqref{regAGP_eqn} in the main text. More explicitly, we need to solve the banded matrix equation
\begin{equation}
\small
\begin{pmatrix}
2+\eta^2-2\alpha_0 & -\rho_0(\alpha_1+1) & -\rho_0\rho_1 & \cdots & 0 \\[3pt]
-\rho_0(\alpha_1+1) & 2+\eta^2+2\alpha_1\alpha_0 & -\rho_1(\alpha_2-\alpha_0) & \cdots & 0 \\[3pt]
-\rho_0\rho_1 & -\rho_1(\alpha_2-\alpha_0) & 2+\eta^2+2\alpha_2\alpha_1 & \ddots & \vdots \\[3pt]
\vdots & \vdots & \ddots & \ddots & -\rho_{M-2}(1-\alpha_{M-3}) \\[3pt]
0 & 0 & \cdots & -\rho_{M-2}(1-\alpha_{M-3}) & 2+\eta^2+2\alpha_{M-2}
\end{pmatrix}
\begin{pmatrix} c_0 \\[3pt] c_1 \\[3pt] c_2 \\[3pt] \vdots \\[3pt] c_{M-1}\end{pmatrix}
=
\mathcal{N}^{-1}\begin{pmatrix} \alpha_0 - 1 \\[3pt] \rho_0\alpha_1 \\[3pt] \rho_0\rho_1 \\[3pt] \vdots \\[3pt] 0 \end{pmatrix} ,
\label{eq:em-matrixeq}
\end{equation}
where the entries in the left-hand side matrix $\mathsf{S}$ are: $\mathsf{S}_{n,n} = 2 + \eta^2 + 2\alpha_n\alpha_{n-1} ,
  \;
  \mathsf{S}_{n,n+1} = -\rho_n\big(\alpha_{n+1}-\alpha_{n-1}\big) ,
  \; \text{and}\,
  \mathsf{S}_{n,n+2} = -\rho_n\rho_{n+1}$, and only the first three terms are non-zero in the right-hand side vector. For the truncation, we set $\alpha_{M-1} = 1$ to keep truncated $\mathcal{U}_\lambda$ unitary (paraorthogonal closure).

\section{Trotterized XXZ chain}
In this section, we discuss the brickwork circuit structure of the Trotterized XXZ chain. We begin by considering a spin-$1/2$ anisotropic Heisenberg (XXZ) chain described by the Hamiltonian
\begin{align}
    H_{XXZ} = J \sum_{j=1}^{L-1} \left(\sigma_j^x\sigma_{j+1}^x + \sigma_j^y\sigma_{j+1}^y + \Delta \;\sigma_j^z\sigma_{j+1}^z\right)\,,
\end{align}
where $\Delta$ controls the anisotropy strength. Setting $\Delta=1$ reduces the model to the isotropic XXX chain, and $\Delta = 0$ corresponds to the free-fermion XX model. The Hamiltonian commutes with the total spin operator $S_z$ and has reflection symmetry for open boundary conditions. The system has an additional spin-flip symmetry at the $S_z=0$ sector. The unitary operator describing the time evolution by the above Hamiltonian can be written as a brickwork circuit, Eq.~\eqref{brickwork_circuit}, by choosing the unitary gates $U_{j,j+1}$ as the $\check{R}$ matrix defined by
\begin{align}
    \check{R} = \begin{pmatrix}
\sin(u+\gamma) & 0 & 0 & 0 \\
0 & \sin\gamma & \sin u & 0 \\
0 & \sin u & \sin\gamma & 0 \\
0 & 0 & 0 & \sin(u+\gamma)
\end{pmatrix}\,, \label{R_matrix}
\end{align}
where $\gamma$ is related to the anisotropy strength as $\Delta = \cos{\gamma}$ and $J = 1/(2\sin \gamma)$, and $u$ is the spectral parameter. This form of Trotterization preserves the integrability of the model (see \cite{Claeys_UnitaryCircuits, Vanicat_IntegrableTrotterization, Doikou_integrability} for details). Unitarity requires that the parameter $u$ be complex. With $u = i\theta,\; \theta \in \mathbb{R}$, and after dividing out the overall factor $\sin(u+\gamma)=\nu$, we can rewrite the $\check{R}$-matrix as
\begin{align}
    \check{R}(\theta,\gamma) = \begin{pmatrix}
1 & 0 & 0 & 0 \\[2pt]
0 & \dfrac{\sin\gamma}{\nu} & \dfrac{i \sinh\theta}{\nu} & 0 \\[6pt]
0 & \dfrac{i \sinh\theta}{\nu} & \dfrac{\sin\gamma}{\nu} & 0 \\[6pt]
0 & 0 & 0 & 1
\end{pmatrix}\,, \label{R_matrix_unitary}
\end{align}
where $\nu(\theta,\gamma) = \sin(\gamma + i\theta) = \sin{\gamma} \cosh{\theta} + i \cos{\gamma}\sinh{\theta}$ is the normalization factor. Since $|\nu|^2 = \sin^2\gamma + \sinh^2\theta$, the inner block is unitary and $\check{R}^\dagger \check{R} = \mathbb{I}$. Unitarity further requires $\gamma \in \mathbb{R}$, hence $|\Delta| = |\cos\gamma| < 1$, so the integrable Trotterization is confined to the gapless (critical) regime of the XXZ chain. The unitary gates with $\check{R}(\theta,\gamma)$ simulates the XXZ chain for small $\theta$ as
\begin{align}
    \lim_{\theta\to0}\, -i(\partial_\theta \check{R})\check{R}^\dagger = \frac{\sigma^x\sigma^x + \sigma^y\sigma^y + \cos{\gamma}\;\sigma^z\sigma^z}{2\sin{\gamma}} - \frac{\cot\gamma}{2}\,\id\,,
\end{align}
which identifies the anisotropy as $\Delta = \cos\gamma$.

We can use $\theta$ as a tunable parameter. We need the Hermitian operator $-i(\partial_\theta U)U^\dagger$ to generate the AGP, which can be calculated using the local seed operators $g_\theta^{\text{loc}}(\gamma) \equiv -i(\partial_\theta \check{R})\check{R}^\dagger\,.$ It can be written in the matrix form as
\begin{align}
    g_\theta^{\text{loc}}(\gamma) = \frac{1}{|\nu|^2}\begin{pmatrix}
0 & 0 & 0 & 0 \\
0 &- \sin\gamma \cos{\gamma}& \sin\gamma \cosh\theta & 0 \\
0 & \sin\gamma \cosh\theta & - \sin\gamma \cos{\gamma} & 0 \\
0 & 0 & 0 & 0
\end{pmatrix}\,, \label{g_loc}
\end{align}
with $|\nu|^2 = \sin^2\gamma + \sinh^2\theta$.  Since each layer of the unitary circuit consists of commuting gates, they have the tensor product structure
\begin{align}
    U_{\text{odd}}(\theta) = \check{R}_{12}(\theta) \otimes \check{R}_{34}(\theta) \otimes \cdots \otimes \check{R}_{L-1,L}(\theta)\,,
\end{align}
in agreement with $U_{\text{odd}} = \prod_j U_{2j-1,2j}$ of Eq.~\eqref{brickwork_circuit}, and likewise for the even layer. We can calculate its derivative as
\begin{align}
    (\partial_\theta U_{\text{odd}})&=
    \sum_{j \in \text{ odd}} \check{R}_{12}(\theta) \otimes \cdots \otimes \partial_\theta\check{R}_{j,j+1}(\theta) \otimes \cdots \nonumber \\
    & =\sum_{j \in \text{ odd}} \check{R}_{12}(\theta) \otimes \cdots \otimes i\, g_\theta ^{\text{loc}}(\gamma)\,\check{R}_{j,j+1}(\theta) \otimes \cdots \,.
\end{align}
We can simplify the expression as $\partial_\theta U_{\text{odd}} = i\,G_{\theta}^{\text{odd}}U_{\text{odd}}$, where $G_{\theta}^{\text{odd}} = \sum_{j \in \text{ odd}} \text{embed}(g_\theta^{\text{loc}},j)$ is the sum of local two-site Hermitian operators. The structure translates similarly to the even layer, using which we can write the full derivative of $U = U_{\text{odd}}U_{\text{even}}$ as
\begin{align}
    -i(\partial_\theta U)U^\dagger = G_{\theta}^{\text{odd}} + U_{\text{odd}}\,G_{\theta}^{\text{even}}\,U^\dagger_{\text{odd}}\,.
\end{align}
We only require the brickwork circuit and the above seed term to formulate the AGP.

\twocolumngrid

\end{document}